\documentclass{amsart}
\usepackage{mathrsfs}
\usepackage{graphicx}
\usepackage{amsmath}
\usepackage{amscd}
\usepackage{amssymb}
\usepackage[all]{xy}

\input xy
\xyoption{all}

\newcommand{\pd}{\partial}

 \newcommand{\bH}{{\mathbb H}}

\newcommand{\bR}{{\mathbb R}}
\newcommand{\bZ}{{\mathbb Z}}

\newcommand{\cH}{{\mathcal H}}

\newcommand{\half}{\frac{1}{2}}

\DeclareMathOperator{\Aut}{Aut}

 \DeclareMathOperator{\tr}{tr}

\newtheorem{thm}{Theorem}[section]
\newtheorem{theorem/definition}{Theorem/Definition}[section]

\theoremstyle{remark}

\theoremstyle{definition}

\newcommand{\be}{\begin{equation}}
\newcommand{\ee}{\end{equation}}
\newcommand{\bea}{\begin{eqnarray}}
\newcommand{\ben}{\begin{eqnarray*}}
\newcommand{\een}{\end{eqnarray*}}
\newcommand{\eea}{\end{eqnarray}}
\newcommand{\bet}{\begin{equation}
\begin{split}}
\newcommand{\eet}{\end{split}
\end{equation}}

\usepackage{color}

\definecolor{yellow}{rgb}{1,1,0}
\definecolor{orange}{rgb}{1,.7,0}
\definecolor{red}{rgb}{1,0,0} \definecolor{green}{rgb}{0,1,1}
\definecolor{white}{rgb}{1,1,1}

\definecolor{A}{rgb}{.75,1,.75}

\newcommand{\corr}[1]{\langle {#1} \rangle}

\newcommand{\Corr}[1]{\biggl\langle {#1} \biggr\rangle}

\begin{document}

\title[Genus Expansions of  Matrix Models: Fat Graphs vs. Thin Graphs ]
{Genus Expansions of Hermitian One-Matrix Models: Fat Graphs vs. Thin Graphs}

\author{Jian Zhou}
\address{Department of Mathematical Sciences\\
Tsinghua University\\Beijing, 100084, China}
\email{jzhou@math.tsinghua.edu.cn}

\begin{abstract}
We consider two different genus expansions
of the free energy functions of Hermitian one-matrix models, one using fat graphs,
one using ordinary graphs (thin graphs).
Some structural results are first proved for the thin version of genus expansion
using renormalized coupling constants,
and then applied to the fat version.
\end{abstract}
\maketitle

\section{Introduction}

This is a sequel to \cite{Zhou-MatMod} where we presented a formula for the $n$-point correlators
in Hermitian one-matrix models.
This result was inspired by a work of Dubrovin-Yang \cite{Dub-Yan} where
a connection between Hermitian one-matrix models and Toda lattice hierarchy
was used.
In \cite{Zhou-MatMod},
a connection to the KP hierarchy  was used instead,
and this enabled us to apply the formula for $n$-point correlations
associated to a $\tau$-function of the KP hierarchy developed
in an earlier work \cite{Zhou-Emergent}.
One of the goals of this paper
is to apply the results on the $n$-point correlations obtained in \cite{Dub-Yan}
and \cite{Zhou-MatMod} to provide some new way to understand about the structure
of the free energy functions of the Hermitian one-matrix models.

To achieve this goal,
we will first distinguish two kinds of genus expansions
for the free energy.
Ever since \cite{tH},
most authors have focused on the genus expansion induced by the genus
of fat graphs.
In this paper,
we will introduce another genus expansion induced by considering
the thin graphs obtained by the skeletons of the fat graphs.
These two kinds of genus expansions will be referred to as
the {\em fat} and {\em thin} genus expansions respectively.

The motivation for introducing the thin genus expansion
comes from another earlier work of the author \cite{Zhou-1D}.
Again influenced by \cite{tH},
a lot of work on matrix models have focused on the large $N$ limits
and in particular double scaling limit of the matrix models to make connections
to topological 2D gravity.
See e.g. the survey \cite{DGZ}.
Going against this direction,
the author considered the case of $N=1$ and referred to the
resulting theory as topological 1D gravity
in \cite{Zhou-1D}.
In that setting,
ordinary graphs instead of fat graphs were used.
More importantly, 
a version of renormaliztion was developed in that theory
and some structural results were proved for the free energy function.
This inspires us to introduce the thin genus expansions
for Hermitian matrix models with finite size $N$ in this work.

The advantage of thin genus expansion against the fat one
will only become clear after we apply the renormalized coupling
constants introduced in \cite{Itzykson-Zuber} and
developed in \cite{Zhou-1D}.
Their definitions will be recalled in \S \ref{sec:Renormalized}.
In the same spirit of \cite{Zhou-1D},
we will prove in two different ways the following main results
of this paper:
The thin free energy of the Hermitian $N \times N$-matrix model
has the following structure:
\bea
&& F_{0,N} = N \sum_{k=0}^\infty  \frac{(-1)^k}{(k+1)!} (I_k+\delta_{k,1}) I_0^{k+1}, \\
&& F_{1,N} = \frac{N^2}{2} \ln \frac{1}{1- I_1},
\eea
and for $g \geq 1$,
\be
F_{g,N}  =  \sum_{\sum_{j=3}^{2g}  (j-2)l_j = 2g-2}
 \corr{p_3^{l_3} \cdots p_{2g}^{l_{2g}}}^c_{g,N}
\prod_{j=3}^{2g} \frac{1}{l_j!j^{l_j}}\biggl( \frac{I_j}{(1-I_1)^{j/2}}\biggr)^{l_j}.
\ee
In particular,
for each $g$,
the computation of $F_{g, N}$ is reduced to finitely many correlators,
hence we get a very effective way to compute them.

Our first proof is based on combining the Virasoro constraints
well-known in the matrix model literature
with the renormalized coupling constants.
By applying the loop equations for Hermitian one-matrix models
to the fat and thin genus expansions,
we get the fat and thin Virasoro constraints respectively.
These constraints provide alternative ways to compute
the fat and thin correlators,
other than the formulas in \cite{Dub-Yan, Zhou-MatMod}.
The first two Virasoro constraints are often called
the {\em puncture equation} and the {\em dilaton equation} in the literature.
As a common practice
they are used to remove the two lowest degree operators in the correlators.
However,
when combined with the renormalized coupling constants,
these two constraints become much more powerful.
As noticed in \cite{Zhou-1D} in the case of topological 1D gravity,
when these constraints are expressed in the normalized coupling constants $\{I_k\}_{k\geq 0}$,
these constraints reduce each thin free energy in genus $g$ to finitely many correlators.
It turns out that the same holds for Hermitian one-matrix models,
and this is one of the ways that we use to prove the above results.

Our second proof is also based on  combining a technique well-known in the matrix model
literature  with the use of renormalized coupling constants.
To evaluate the Gaussian integral on the space of Hermitian $N\times N$-matrices,
one can reduce it to an integral on $\bR^N$.
One can then apply the change of coupling constants trick developed in \cite{Zhou-1D}
for formal Gaussian integrals on $\bR^1$.

Both of the above proofs cease to work for the fat genus expansion,
nevertheless,
similar results still hold but now infinitely many fat correlators are
involved at each genus.
The reason is that the fat and thin correlators satisfy different selection rules,
and the fat selection rule no longer exclude enough fat correlators.
An alternative way to get results for the fat genus expansion is to apply the results
for the thin genus expansion.
This will be explained in \S \ref{sec:Fat}.

The rest of the paper is arranged as follows.
In \S \ref{sec:Expansion} we define two kinds of genus expansions
of the free energies of Hermitian one-matrix models.
We then recall the derivation of Virasoro constraints and specialize
them to the two genus expansion.
After we apply the thin Virasoro constraints to obtain
some general results on the thin genus $g$ free energy $F_{g,N}$
in \S \ref{sec:More-App},
we prove our main results on $F_{g, N}$ in \S \ref{sec:Renormalized}
and \S \ref{sec:Renormalization} in two different ways.
In the final \S \ref{sec:Fat}
we derive similar results for the fat free energy.

\section{Two Kinds of Genus Expansions
in Hermitian Matrix Models}

\label{sec:Expansion}

In this Section we introduce two kinds of genus expansions for the free energy
functions of Hermitian one-matrix models.
We also present the fat and thin selection rules.
For general references on matrix models,
see \cite{Mehta, BIZ, BKSWZ}.

\subsection{Hermitian one-matrix models}

For each $N$, the partition function of the Hermitian $N\times N$-matrix model is
defined by the formal Gaussian integral:
\be
Z_N = \frac{\int_{\bH_N} dM \exp \biggl( \tr
\sum\limits_{n=1}^\infty \frac{g_n-\delta_{n,2}}{ng_s} M^n\biggr)}{\int_{\bH_N} dM
\exp \biggl( -\frac{1}{2g_s} \tr (M^2)\biggr)},
\ee
where $\bH_N$ is the space of Hermitian $N \times N$-matrices.
Its {\em free energy} $F_N$ is defined by:
\be
F_N : = \log Z_N.
\ee
The first few terms of $F_N$ are given by:
\ben
F_N & = & \frac{1}{2}N^2g_2+\frac{1}{2}Ng_s^{-1}g_1^2
+(\frac{1}{2}N^3+\frac{1}{4}N)g_s g_4+N^2g_3g_1+\frac{N^2}{4}g_2^2
+ \frac{N}{2}g_s^{-1} g_2g_1^2 \\
& + & (\frac{5N^2}{3}+\frac{5N^4}{6})g_s^2g_6
 +(N+2N^3)g_sg_5g_1 + (\frac{N}{2}+N^3) g_s g_4g_2 + \frac{3N^2}{2} g_4g_1^2 \\
& + & \biggl(\frac{N}{6} +\frac{2N^3}{3} \biggr) g_s g_3^2
+  2N^2 g_3g_2g_1 + \frac{N}{3}g_s^{-1}g_3g_1^3
+\frac{N^2}{6} g_2^3 + \frac{N}{2} g_s^{-1} g_2^2g_1^2 + \cdots
\een
For a partition $\lambda = (\lambda_1, \dots, \lambda_l)$,
the correlator $\corr{\frac{1}{z_\lambda} p_\lambda}_N^c$
is defined by
\be
\corr{\frac{1}{z_\lambda} p_\lambda}_N^c:
= \frac{\pd^l F_N}{\pd g_{\lambda_1} \cdots \pd g_{\lambda_l}}\biggl|_{g_i = 0, i =1, \dots},
\ee
where $p_\lambda = p_{\lambda_1}\cdots p_{\lambda_l}$.
Here we are following the notations of \cite{Macdonald}.
By using the fat graphs introduced in \cite{tH},
one can get (see e.g. \cite[(21)]{Zhou-MatMod}:
\be \label{eqn:correlator}
\corr{\frac{1}{z_\lambda} p_\lambda}_N^c
= \sum_{\Gamma \in \Gamma^{\lambda c}}
\frac{1}{|\Aut(\Gamma)|} g_s^{\half|\lambda| - l(\lambda)} N^{|F(\Gamma)|},
\ee
where $\Gamma^{\lambda c}$ is the set of connected fat graphs of type $\lambda$.
The free energy function $F_N$ is given by these correlators as follows:
\be
F_N = \sum_\lambda \corr{\frac{1}{z_\lambda} p_\lambda}_N^c g_\lambda.
\ee

\subsection{Genus expansion by thin graphs} \label{sec:Thin}

The first kind of genus expansion of $F_N$  we consider is of the following form:
\ben
F_N = \sum_{g\geq 0} g_s^{g-1} F_{g, N}.
\een
We will refer to it as the {\em thin genus expansion} of $F_N$.
For example,
\ben
F_{0,N} & = & \frac{1}{2}N g_1^2
+ \frac{N}{2} g_2g_1^2
+ \frac{N}{3} g_3g_1^3 + \frac{N}{2} g_2^2g_1^2
+ \frac{N}{2} g_2^3g_1^2 + Ng_2g_3g_1^3 + \frac{N}{4} g_4g_1^4 + \cdots \\
F_{1,N} & = & \frac{1}{2}N^2g_2
 +N^2g_3g_1+\frac{N^2}{4}g_2^2    + \frac{3N^2}{2} g_4g_1^2
+  2N^2 g_3g_2g_1 +\frac{N^2}{6} g_2^3   + \cdots, \\
F_{2,N} & = &
(\frac{1}{2}N^3+\frac{1}{4}N) g_4
+ (N+2N^3)g_5g_1 + (\frac{N}{2}+N^3) g_4g_2  \\
& + & \biggl(\frac{N}{6} +\frac{2N^3}{3} \biggr) g_3^2
+ \cdots, \\
F_{3,N} & = & (\frac{5N^2}{3}+\frac{5N^4}{6}) g_6
+ \cdots.
\een

\subsection{Thin correlators and thin selection rule}

Suppose that the correlator $\corr{p_{a_1} \cdots p_{a_n}}^c_N$
contributes to $F_{g,N}$.
It is a summation over thin graphs $\hat{\Gamma}$, such that
the number of vertices $V(\hat{\Gamma}) = n$,
the number of edges $E(\hat{\Gamma}) = \half \sum_{j=1}^n a_j$.
The number of loops of the thin graph $\hat{\Gamma}$ is
\be
g(\hat{\Gamma}) = V(\hat{\Gamma}) - E(\hat{\Gamma}) + 1 = n - \half \sum_{j=1}^n a_j +1.
\ee
Therefore,
$\corr{p_{a_1} \cdots p_{a_n}}^c_N$
contributes to $F_{g,N}$ iff
\be \label{eqn:Thin-Select}
\sum_{j=1}^n a_j = 2g-2 +2n.
\ee
This will be referred to as the {\em thin selection rule}.
We will write  $\corr{p_{a_1} \cdots p_{a_n}}^c_{g,N}$
instead of $\corr{p_{a_1} \cdots p_{a_n}}^c_N$ when this selection rule is satisfied
for genus $g$.
It will be referred to as a genus $g$ thin correlator.

For example,
in degree two we have
\begin{align*}
\corr{\frac{p_2}{2}}^c_{1,N} & = \half N^2,
& \corr{p_1^2}_{0,N}^c & = N ,
\end{align*}
in degree four,
\begin{align*}
\corr{\frac{p_4}{4}}_{2,N}^c & = \frac{1}{2}N^3 + \frac{1}{4} N, &
\corr{\frac{p_3}{3}p_1}_{1,N}^c &= N^2, \\
\corr{(\frac{p_2}{2})^2}_{1,N}^c & = \frac{1}{2} N^2, &
\corr{\frac{p_2}{2}p_1^2}_{0,N}^c &= N,  \\
\corr{p_1^4}_{g,N}^c & = 0, \qquad g \geq 0, & &
\end{align*}
and in degree $6$:
\begin{align*}
\corr{\frac{p_6}{6}}_{3,N}^c & = \frac{5N^2}{3}+\frac{5N^4}{6},
& \corr{\frac{p_5}{5}p_1}_{2,N}^c &  = N+2N^3, \\
\corr{\frac{p_4}{4}\frac{p_2}{2}}_{2,N}^c & = \frac{N}{2}+N^3,
& \corr{\frac{p_4}{4}p_1^2}_{1,N}^c & = 3N^2, \\
\corr{(\frac{p_3}{3})^2}_{2,N}^c & = \frac{N}{3} +\frac{4N^3}{3},
& \corr{\frac{p_3}{3}\frac{p_2}{2}p_1}_{1,N}^c & = 2N^2, \\
\corr{\frac{p_3}{3}p_1^3}^c_{0,N} & = 2N,  &
\corr{(\frac{p_2}{2})^3}_{1,N}^c & = N^2, \\
\corr{(\frac{p_2}{2})^2p_1^2}_{0,N}^c &= 2N, &
\corr{\frac{p_2}{2} p_1^4}_{g, N}^c & = 0, \qquad g \geq 0, \\
\corr{p_1^6}_{g, N}^c & = 0, \qquad g \geq 0.
\end{align*}

\subsection{The case of $N=1$}

As mentioned in the Introduction,
the motivation of considering the thin-graph expansion comes from the $N=1$ case
of the $N \times N$-matrix model  studied in \cite{Zhou-1D}.
Its partition function is defined by the formal Gaussian integral:
\be
Z^{1D} = \frac{1}{\sqrt{2\pi}\lambda}
\int  dx \exp \frac{1}{\lambda^2} \biggl( -\half x^2 + \sum_{n \geq 1} t_{n-1} \frac{x^n}{n!}  \biggr).
\ee
This is just $Z_{N=1}$ with
\bea
g_s & = & \lambda^2, \\
g_n & = & (n-1)! t_{n-1}, \qquad n \geq 1.
\eea
The first few terms of the free energy $F^{1D}= \log Z^{1D}$ are given by:
\ben
F^{1D} & = & (\frac{1}{2}\lambda^{-2}t_0^2+\frac{1}{2}t_1)
+ (\frac{1}{2}t_0^2t_1\lambda^{-2} +\frac{1}{2}t_0t_2+\frac{1}{4}t_1^2 + \frac{1}{8} t_3\lambda^2) \\
& + & (\frac{1}{2}t_0^2t_1^2\lambda^{-2}+\frac{1}{6} t_0^3t_2\lambda^{-2}
+ \frac{1}{6}t_1^3 + \frac{1}{4}t_0^2t_3+t_0t_2t_1 \\
& + & \frac{5}{24}t_2^2\lambda^2
+ \frac{1}{8}t_0t_4\lambda^2+\frac{1}{4}t_1t_3\lambda^2 +\frac{1}{48}t_5\lambda^4) \\
& + & (\frac{1}{2}\lambda^{-2}t_1^3t_0^2 + \frac{1}{2}\lambda^{-2}t_1t_2t_0^3 + \frac{1}{24}\lambda^{-2}t_3t_0^4 \\
& + & \frac{1}{8} t_1^4  + \frac{3}{4} t_0^2t_1t_3 + \frac{3}{2} t_0 t_1^2t_2 + \frac{1}{12} t_0^3 t_4
+ \frac{1}{2} t_0^2t_2^2 \\
& + & \frac{5}{8} \lambda^2 t_2^2t_1 + \frac{3}{8} \lambda^2t_4t_0t_1 + \frac{3}{8} \lambda^2t_3t_1^2
+ \frac{1}{16} \lambda^2t_5t_0^2   + \frac{2}{3} \lambda^2t_3t_0t_2 \\
& + & \frac{1}{16}\lambda^4t_5t_1
+ \frac{1}{12} \lambda^4t_3^2 + \frac{1}{48} \lambda^4t_0t_6
+ \frac{7}{48} \lambda^4t_4t_2 + \frac{1}{384} t_7\lambda^6 ) + \cdots,
\een
and one has
\ben
F_0^{1D} & = & \frac{1}{2} t_0^2
+ \frac{1}{2}t_0^2t_1
 +  \frac{1}{2}t_0^2t_1^2 +\frac{1}{6} t_0^3t_2
+ \frac{1}{2} t_1^3t_0^2 + \frac{1}{2} t_1t_2t_0^3 + \frac{1}{24} t_3t_0^4
+ \cdots,
\een
\ben
F_1^{1D} & = & \frac{1}{2}t_1
+ \frac{1}{2}t_0t_2+\frac{1}{4}t_1^2
+ \frac{1}{6}t_1^3 + \frac{1}{4}t_0^2t_3+t_0t_2t_1 \\
& + & \frac{1}{8} t_1^4  + \frac{3}{4} t_0^2t_1t_3
+ \frac{3}{2} t_0 t_1^2t_2 + \frac{1}{12} t_0^3 t_4
+ \frac{1}{2} t_0^2t_2^2   + \cdots,
\een
\ben
F_2^{1D} & = & \frac{1}{8} t_3
+ \frac{5}{24}t_2^2
+ \frac{1}{8}t_0t_4+\frac{1}{4}t_1t_3 \\
& + & \frac{5}{8} t_2^2t_1 + \frac{3}{8} t_4t_0t_1
+ \frac{3}{8} t_3t_1^2
+ \frac{1}{16} t_5t_0^2   + \frac{2}{3} t_3t_0t_2
+ \cdots,
\een
\ben
F_3^{1D} & = &  \frac{1}{48}t_5
+ \frac{1}{16}t_5t_1
+ \frac{1}{12} t_3^2 + \frac{1}{48} t_0t_6
+ \frac{7}{48} t_4t_2 + \cdots,
\een
\ben
F_4^{1D} & = &   \frac{1}{384} t_7+ \cdots.
\een
One can check that they match with the first few terms of $F_{g, N=1}$.

\subsection{Genus expansion by fat graphs}

Another way to define a genus expansion of $F_N$ is to
introduce the  't Hooft coupling constant£º
\be
t = N g_s.
\ee
With this one can substitute $N$ by $tg_s^{-1}$ in $F_N$ to get:
\ben
F_N & = & \frac{1}{2}t^2g_s^{-2}g_2+\frac{1}{2}tg_s^{-2}g_1^2
+(\frac{1}{2}t^3g_s^{-2}+\frac{1}{4}t) g_4
+t^2g_s^{-2}g_3g_1 \\
& + & \frac{t^2}{4}g_s^{-2}g_2^2
+ \frac{t}{2}g_s^{-2} g_2g_1^2
+ (\frac{5t^2}{3}+\frac{5t^4}{6}g_s^{-2})g_6
 +(t+2t^3g_s^{-2})g_5g_1 \\
& + & (\frac{t}{2}+t^3 g_s^{-2}) g_4g_2
 + \frac{3t^2}{2} g_s^{-2}g_4g_1^2
+ \biggl(\frac{t}{6} +\frac{2t^3}{3} g_s^{-2} \biggr)  g_3^2
+  2t^2 g_s^{-2}g_3g_2g_1 \\
& + & \frac{t}{3}g_s^{-2}g_3g_1^3
+ \frac{t^2}{6} g_s^{-2} g_2^3 + \frac{t}{2} g_s^{-2} g_2^2g_1^2 + \cdots.
\een
We will write
\be
F_N = \sum_{g \geq 0}g_s^{2g-2} F_g(t),
\ee
and refer to it as the {\em fat genus expansion}.
For example,
 \ben
F_{0}(t) & = & \frac{1}{2}t^2g_2+\frac{1}{2}tg_1^2
+\frac{1}{2}t^3g_4
+t^2g_3g_1+\frac{t^2}{4}g_2^2
+ \frac{t}{2} g_2g_1^2 \\
& + & \frac{5t^4}{6}g_6
 +2t^3g_5g_1 + t^3 g_4g_2
 + \frac{3t^2}{2} g_4g_1^2 \\
& + & \frac{2t^3}{3}  g_3^2
+  2t^2 g_3g_2g_1 + \frac{t}{3}g_3g_1^3
+\frac{t^2}{6} g_2^3 + \frac{t}{2} g_2^2g_1^2 + \cdots \\
& = & t( \frac{1}{2}g_1^2
+ \frac{1}{2} g_2g_1^2 + \frac{t}{3}g_3g_1^3
+ \frac{1}{2} g_2^2g_1^2 + \cdots) \\
& + & t^2(\frac{1}{2}g_2 +g_3g_1+\frac{1}{4}g_2^2
 + \frac{3}{2} g_4g_1^2 + 2g_3g_2g_1
+\frac{1}{6} g_2^3 + \cdots) \\
& + & t^3(\frac{1}{2}g_4
 +2 g_5g_1 + g_4g_2
+ \frac{2}{3}  g_3^2  + \cdots) \\
& + & t^4 (  \frac{5}{6}g_6  + \cdots) + \cdots,
\een
\ben
F_{1}(t) & = &   \frac{1}{4}t g_4
+ \frac{5t^2}{3} g_6 +t g_5g_1 + \frac{t}{2} g_4g_2
+ \frac{t}{6}   g_3^2 + \cdots.
\een
It is clear that $F_g(t)$ is a formal power series in $t$:
\be
F_g(t) = \sum_{m \geq 1} F_{g, m} t^m.
\ee

\subsection{Fat correlators and the fat selection rule}

The fat correlators are defined by:
\be
\corr{\frac{1}{z_\lambda} p_\lambda}_g^c(t):
= \frac{\pd^l F_g(t)}{\pd g_{\lambda_1} \cdots \pd g_{\lambda_l}}\biggl|_{g_i = 0, i =1, \dots},
\ee
By \eqref{eqn:correlator} we have
\be \label{eqn:correlator2}
\begin{split}
\corr{\frac{1}{z_\lambda} p_\lambda}_N^c
& = \sum_{\Gamma \in \Gamma^{\lambda c}}
\frac{1}{|\Aut(\Gamma)|} g_s^{\half|\lambda| - l(\lambda)-|F(\Gamma)|} t^{|F(\Gamma)|} \\
& = \sum_g g_s^{2g-2}\sum_{\Gamma \in \Gamma_g^{\lambda c}}
\frac{1}{|\Aut(\Gamma)|}  t^{|F(\Gamma)|},
\end{split}
\ee
where $\Gamma^{\lambda c}_g$ is the set of connected fat graphs of type $\lambda$
and of genus $g$.
It follows that
\be
\corr{\frac{1}{z_\lambda} p_\lambda}_g^c(t)
= \sum_{\Gamma \in \Gamma_g^{\lambda c}}
\frac{1}{|\Aut(\Gamma)|}
t^{|F(\Gamma)|},
\ee
and so $\corr{\frac{1}{z_\lambda} p_\lambda}_g^c(t) \neq 0$ only if
\be
2g-2 = \half|\lambda| - l(\lambda) - m
\ee
for some $m \geq 1$.
In other words,
a fat correlator $\corr{p_{a_1}\cdots p_{a_n}}^c_{\tilde{g}}(t)$ is nonzero
only when
\be \label{eqn:Fat-Select}
\sum_{i=1}^n a_i = 4\tilde{g}- 4+ 2n +2m
\ee
for some $m \geq 1$.
We will refer to this as the {\em fat selection rule}.
By comparing with the thin selection rule,
we see that when the thin correlator $\corr{p_{a_1}\cdots p_{a_n}}^c_{g, N}$
and the fat correlator $\corr{p_{a_1}\cdots p_{a_n}}^c_{\tilde{g}}(t)$
are both nonzero,
\be
g = 2\tilde{g} + m -1
\ee
for some $m \geq 1$.

The following are some examples of fat correlators.
In degree two we have
\begin{align*}
\corr{\frac{p_2}{2}}^c_{0}(t) & = \half t^2,
& \corr{p_1^2}_{0}^c(t) & = t,
\end{align*}
in degree four,
\begin{align*}
\corr{\frac{p_4}{4}}_{0}^c(t) & = \frac{1}{2}t^3, & 
\corr{\frac{p_4}{4}}_{1}^c(t) & = \frac{1}{4} t, &
\corr{\frac{p_3}{3}p_1}_{0}^c(t) &= t^2, \\
\corr{(\frac{p_2}{2})^2}_{0}^c(t) & = \frac{1}{2} t^2, &
\corr{\frac{p_2}{2}p_1^2}_{0}^c(t) &= t,  &
\corr{p_1^4}_{g}^c(t) & = 0, \qquad g \geq 0, 
\end{align*}
and in degree $6$:
\begin{align*}
\corr{\frac{p_6}{6}}_{0}^c(t) & = \frac{5t^4}{6}, &
\corr{\frac{p_6}{6}}_{1}^c(t) & = \frac{5t^2}{3}, & 
\corr{\frac{p_5}{5}p_1}_{0}^c(t) &  = 2t^3, \\
\corr{\frac{p_5}{5}p_1}_{1}^c(t) &  = t, &
\corr{\frac{p_4}{4}\frac{p_2}{2}}_{0}^c(t) & = t^3, &
\corr{\frac{p_4}{4}\frac{p_2}{2}}_{1}^c(t) & = \frac{t}{2}, \\
\corr{\frac{p_4}{4}p_1^2}_{0}^c(t) & = 3t^2, &
\corr{(\frac{p_3}{3})^2}_{0}^c(t) & = \frac{4t^3}{3}, &
\corr{(\frac{p_3}{3})^2}_{1}^c(t) & = \frac{t}{3}, \\
\corr{\frac{p_3}{3}\frac{p_2}{2}p_1}_{0}^c(t) & = 2t^2, &
\corr{\frac{p_3}{3}p_1^3}^c_{0}(t) & = 2t,  &
\corr{(\frac{p_2}{2})^3}_{0}^c(t) & = t^2, \\
\corr{(\frac{p_2}{2})^2p_1^2}_{0}^c(t) &= 2t, &
\corr{\frac{p_2}{2} p_1^4}_{g, N}^c & = 0, \qquad g \geq 0, &
\corr{p_1^6}_{g, N}^c & = 0, \qquad g \geq 0.
\end{align*}

\section{Virasoro Constraints}

\label{sec:Virasoro}

We recall the derivation of Virasoro constraints for matrix models for finite $N$
in the literature.
We specialize them to the two genus expansions discussed above.

\subsection{Loop operator and loop equations}

We now recall the derivation of loop equations in Hermitian matrix models.
See e.g. Kazakov's contribution to \cite{BKSWZ}.
For simplicity of notations, rewrite $Z_N$ as follows:
\be
Z_N = \frac{\int_{\bH_N} dM \exp \biggl( \tr
\sum\limits_{n=1}^\infty \tilde{T}_n M^n\biggr)}
{\int_{\bH_N} dM \exp \biggl( -\frac{1}{2g_s} \tr (M^2)\biggr)}.
\ee
where $\tilde{T}_n = \frac{g_n-\delta_{n,2}}{ng_s}$.
It can be converted to a formal integral over $\bR^N$
(see e.g. \cite{BIZ}):
\be \label{eqn:Log-gas}
Z_N= c_N\int_{\bR^N} \prod^N_{i=1} d\lambda_i \cdot
\prod^N_{i=1}\exp \biggl(
\sum_{n=1}^\infty \tilde{T}_n \lambda^n_i \biggr) \cdot \prod_{1\leq i<j \leq N}
(\lambda_i - \lambda_j)^2,
\ee
where $C_N$ is a constant depending on $N$.

Consider the collective {\em loop operator}:
\be
W_N(z) =
\sum_{k=1}^N \frac{1}{z - \lambda_k} = \tr \biggl(\frac{1}{z -M} \biggr).
\ee
Start with the identity:
\ben
&& \int_{\bR^N} \prod^N_{i=1} d\lambda_i \cdot \sum_{k=1}^N
\frac{\pd}{\pd \lambda_k}
\biggl( \frac{1}{z-\lambda_k} \prod^N_{i=1} \exp \biggl(
\sum_{n=0}^\infty
\tilde{T}_n\lambda^n_i \biggr) \cdot \prod_{1\leq i<j \leq N}
(\lambda_i - \lambda_j)^2  \biggr)
= 0.
\een
Rewrite the left-hand side as follows:
\ben
&& \Corr{\sum_{k=1}^N \biggl(\frac{1}{(z-\lambda_k)^2}
+ \frac{1}{z-\lambda_k} \cdot \biggl(\sum_{n=0}^\infty n\tilde{T}_n \lambda_k^{n-1}
+ 2 \sum_{j \neq k} \frac{1}{\lambda_k - \lambda_j}
\biggr) \biggr)}_{N,T} = 0.
\een
Using the identity
\be
\sum_{k=1}^N \frac{1}{(z - \lambda_k)^2} + 2  \sum_{1 \leq j \neq k \leq N}
\frac{1}{z -\lambda_k}\frac{1}{\lambda_k-\lambda_j}
= \biggl(\sum_{k=1}^N \frac{1}{z-\lambda_k} \biggr)^2,
\ee
one finds:
\be
\Corr{W_N^2(z) +
\sum_{k=1}^N \frac{1}{z -\lambda_k}
\sum_{n\geq 0} n\tilde{T}_n \lambda_k^{n-1}}_{N,T} = 0.
\ee
The summation over $k$ can be reexpressed as a residue:
\be \label{eqn:Loop}
\Corr{W_N(z)^2+ \oint_C \frac{d\tilde{z}}{2\pi i}
i_{z, \tilde{z}}\frac{1}{z-\tilde{z}} \cdot W_N(\tilde{z})\sum_{n\geq 1} n \tilde{T}_n \tilde{z}^{n-1}
}_{N,T} = 0,
\ee
where $C$ is a large enough circle,
and
\ben
i_{z, \tilde{z}}\frac{1}{z-\tilde{z}} = \sum_{n \geq 0} \frac{\tilde{z}^n}{z^{n+1}}.
\een
This is called the {\em loop equation}.

\subsection{Reformulation in terms of a bosonic field}

The loop equation can be further reformulated by
introducing the collective field
\bea
\Phi_N(z) & = &  \frac{1}{\sqrt{2}} \sum_{n\geq 1} \tilde{T}_nz^n
- \sqrt{2}\tr \log \biggl( \frac{1}{z - M} \biggr) \\
& = & \frac{1}{\sqrt{2}} \sum_{n\geq 1} \tilde{T}_nz^n
+ \sqrt{2} N \log z
- \sqrt{2} \sum_{n\geq 1} \frac{z^{-n}}{n} \frac{\pd}{\pd T_n}.
\eea
The second line follows from the fact
that the insertion of the operator $\tr(M^n) = \sum_{i=1}^N \lambda_i^N$
can be realized by taking a partial
derivative with respect to $T_n$.
Similarly,
\be
W_N(z) = \frac{N}{z} + \sum_{n\geq 1} \frac{1}{z^{n+1}} \frac{\pd}{\pd T_n}.
\ee
Note
\ben
\pd_z \Phi_N(z)
& = & \frac{1}{\sqrt{2}} \sum_{n \geq 1} n \tilde{T}_n z^{n-1}
+\sqrt{2} \biggl(\frac{N}{z} +\sum_{n\geq 1}z^{-n-1} \frac{\pd}{\pd T_n}\biggr) \\
& = & \frac{1}{\sqrt{2}} \sum_{n \geq 1} n \tilde{T}_n z^{n-1} + \sqrt{2} W_N(z).
\een
The loop equation \eqref{eqn:Loop} can now be rewritten as
\be
\oint_C  \frac{d\tilde{z}}{2 \pi i}
i_{z, \tilde{z}} \frac{1}{z-\tilde{z}}  \Corr{(\pd_{\tilde{z}} \Phi_N(\tilde{z}))^2}_{N,T} = 0,
\ee
or
\be
\oint_C \frac{d\tilde{z}}{ 2\pi i} i_{z, \tilde{z}} \frac{1}{z-\tilde{z}} \cdot T_N(\tilde{z})
Z_N[T] = 0,
\ee
where $T_N(z)$ is the energy-momentum defined by:
\be
T_N(z) = \frac{1}{2} : (\pd_z\Phi_N(z))^2 :
\ee

\subsection{Virasoro constraints}

Expand $T_N(z)$ in the following form:
\be
T_N(z) := \sum_{n \in \bZ} L_{n,N} z^{-n-2},
\ee
where
\ben
&& L_{-1,N} = \sum_{n \geq 1} (n+1)\tilde{T}_{n+1} \frac{\pd}{\pd T_n}
+ NT_1, \\
&& L_{0,N} = \sum_{n \geq 1} n\tilde{T}_n \frac{\pd}{\pd T_n} + N^2, \\
&& L_{1,N} = \sum_{n \geq 1} n\tilde{T}_n \frac{\pd}{\pd T_{n+1}}
+ 2 N \frac{\pd}{\pd T_1}, \\
&& L_{n,N} = \sum_{k \geq 1} k\tilde{T}_k \frac{\pd}{\pd T_{k+n}}
+ \sum_{k=1}^{n} \frac{\pd}{\pd T_k} \frac{\pd}{\pd T_{n-k}}
+ 2 N \frac{\pd}{\pd T_n}, \qquad n \geq 2.
\een
The loop equation can be rewritten as a set of linear differential equations
\be
L_{n,N} Z_N[t] = 0 \;\;\; (n \geq -1).
\ee
These are called {\em Virasoro constraints} because
\be
[L_{m,N}, L_{n,N}] = (m-n) L_{m+n, N}.
\ee

\subsection{Virasoro constraints for thin genus expansion}

Now if we take $\tilde{T}_n = \frac{g_n-\delta_{n,2}}{ng_s}$,
then the operators $L_{n,N}$ become:
\ben
&& L_{-1,N} = - \frac{\pd}{\pd g_1}
+ \sum_{n \geq 1} ng_{n+1} \frac{\pd}{\pd g_n} + Ng_1g_s^{-1}, \\
&& L_{0,N} = - 2\frac{\pd}{\pd g_2} + \sum_{n \geq 1} ng_n \frac{\pd}{\pd g_n}
+ N^2, \\
&& L_{1,N} = - 3\frac{\pd}{\pd g_3} + \sum_{n \geq 1} (n+1)g_n \frac{\pd}{\pd g_{n+1}}
+ 2Ng_s\frac{\pd}{\pd g_1}, \\
&& L_{m,N} = \sum_{k \geq 1} (k+m) (g_k - \delta_{k,2} ) \frac{\pd}{\pd g_{k+m}} +
g_s^2 \sum_{k=1}^{m-1} k(m-k)\frac{\pd}{\pd g_k} \frac{\pd}{\pd g_{m-k}} \\
&& \qquad\qquad + 2 Nm g_s \frac{\pd}{\pd g_m}, \qquad m \geq 2.
\een
We will refer to the constraints 
\be
L_{m, N} Z_N = 0
\ee
as the {\em thin Virasoro constraints}.

\subsection{Thin Virasoro constraints in terms of thin correlators}
It is useful for practical computations to
rewrite the thin Virasoro constraints in terms of thin correllators.
The thin puncture equation can be written as
\be
\corr{p_1 \cdot \frac{p_{a_1}}{a_1} \cdots \frac{p_{a_n}}{a_n}}^c_{g, N} =
\sum_{j=1}^n (a_j-1) \cdot \corr{\frac{p_{a_1}}{a_1} \cdots
\frac{p_{a_j-1}}{a_j-1} \cdots \frac{p_{a_n}}{a_n}}_{g, N}^c,
\ee
together with initial value:
\be
\corr{p_1^2}_{0,N}^c = N
\ee
The thin dilaton equation can be written as
\be
\corr{p_2 \cdot \frac{p_{a_1}}{a_1} \cdots \frac{p_{a_n}}{a_n}}^c_{g, N} =
\sum_{j=1}^n a_j \cdot \corr{\frac{p_{a_1}}{a_1} \cdots \frac{p_{a_n}}{a_n}}_{g, N}^c,
\ee
together with initial value:
\be
\corr{p_2}_{1,N}^c = N^2.
\ee
The third equation in the sequence can be written  as
\be
\begin{split}
\corr{p_3 \cdot \frac{p_{a_1}}{a_1} \cdots \frac{p_{a_n}}{a_n}}^c_{g, N}
& =\sum_{j=1}^n (a_j+1) \cdot \corr{\frac{p_{a_1}}{a_1} \cdots
\frac{p_{a_j+1}}{a_j+1} \cdots \frac{p_{a_n}}{a_n}}_{g, N}^c \\
& + 2N \corr{p_1 \cdot \frac{p_{a_1}}{a_1} \cdots \frac{p_{a_n}}{a_n}}^c_{g-1, N} ,
\end{split}
\ee
and for $m \geq 2$
\be
\begin{split}
& \corr{p_ {m+2} \cdot \frac{p_{a_1}}{a_1} \cdots \frac{p_{a_n}}{a_n}}^c_{g, N}
=\sum_{j=1}^n (a_j+m) \cdot \corr{\frac{p_{a_1}}{a_1} \cdots
\frac{p_{a_j+m}}{a_j+m} \cdots \frac{p_{a_n}}{a_n}}_{g, N}^c \\
& \qquad + 2N \corr{p_m \cdot \frac{p_{a_1}}{a_1} \cdots \frac{p_{a_n}}{a_n}}^c_{g-1, N} \\
& \qquad + \sum_{k=1}^m \corr{p_kp_{m-k} \cdot \frac{p_{a_1}}{a_1} \cdots \frac{p_{a_n}}{a_n}}^c_{g-2, N} \\
& \qquad + \sum_{k=1}^m \sum_{\substack{g_1+g_2=g-1\\I_1 \coprod I_2 = [n]}}
\corr{p_k \cdot \prod_{i\in I_1} \frac{p_{a_i}}{a_i}}^c_{g_1, N}
\cdot \corr{p_{m-k} \cdot \prod_{i\in I_2} \frac{p_{a_i}}{a_i}}^c_{g_2, N},
\end{split}
\ee
where $[n]=\{1, \dots, n\}$.

Now we present some examples of computations of thin correlators by the thin Virasoro constraints
and compare with the results in \cite{Dub-Yan}.
In degree $8$ we have:
\ben
\corr{\frac{p_3}{3}\frac{p_5}{5}}_{3, N}^c
& = &\frac{1}{3} \cdot \biggl( 6 \corr{\frac{p_6}{6}}_{3,N}^c
+ 2N \corr{p_1\frac{p_5}{5}}_{2,N}^c\biggr) \\
& = & \frac{1}{3} \cdot \biggl( 6 \cdot \big(\frac{5N^2}{3} +\frac{5N^4}{6}\big)
+ 2N \cdot (N+2N^3) \biggr) \\
& = & 4N^2 + 3N^4.
\een
Here is another example in degree $8$:
\ben
\corr{(\frac{p_4}{4})^2}_{3, N}^c
& = &\frac{1}{4} \cdot \biggl( 6 \corr{\frac{p_6}{6}}_{3,N}^c
+ 2N \corr{p_2\frac{p_4}{4}}_{2,N}^c + \corr{p_1p_1 \frac{p_4}{4}}_{1,N}^c \biggr) \\
& = & \frac{1}{4} \cdot \biggl( 6 \cdot \big(\frac{5N^2}{3} +\frac{5N^4}{6}\big)
+ 2N \cdot 2(\frac{N}{2}+N^3) + 3N^2 \biggr) \\
& = & \frac{1}{4}(15N^2 + 9N^4).
\een
This matches with \cite[Example 3.2.5]{Dub-Yan}.
Here is an example in degree $10$:
\ben
\corr{(\frac{p_3}{3})^2\frac{p_4}{4} }_{3,N}^c
& = & \frac{1}{3}\biggl(4 \cdot \corr{\frac{p_4}{4}\frac{p_4}{4}}_{3,N}^c
+ 5 \cdot \corr{\frac{p_3}{3} \frac{p_5}{5}}_{3,N}^c
+ 2N \corr{p_1 \frac{p_3}{3}\frac{p_4}{4}}_{2, N}^c \biggr) \\
& = & \frac{1}{3}\biggl(4\cdot \frac{1}{4}(15N^2+9N^4 ) +5(4N^2+3N^4) +2N(2N+6N^3) \biggr) \\
& = & 13N^2 +12N^4.
\een
This matches with \cite[Example 3.2.6]{Dub-Yan}.
In the above we have used the following computation:
\ben
\corr{p_1 \frac{p_3}{3}\frac{p_4}{4}}_{2, N}^c
& = & 2\corr{\frac{p_2}{2}\frac{p_4}{4}}_{2, N}^c
+ 3 \corr{\frac{p_3}{3}\frac{p_3}{3}}_{2, N}^c \\
& = & 2 \cdot (\frac{N}{2} +N^3) +3 \cdot (\frac{N}{3} +\frac{4N^3}{3}) \\
& = & 2N + 6N^3.
\een
Finally,
we present an example in degree $12$:
\ben
\corr{(\frac{p_3}{3})^4}_{3,N}^c
& = & \frac{1}{3}\biggl( 3 \cdot 4\cdot\corr{(\frac{p_3}{3})^2\frac{p_4}{4}}_{3,N}^c
+ 2N \corr{p_1 (\frac{p_3}{3})^3}_{2,N}^c\biggr) \\
& = & \frac{1}{3}\biggl(12 (13N^2+12N^4) + 2N(6N+24N^3) \biggr) \\
& = & 56N^2+64N^4,
\een
where we have used:
\ben
\corr{p_1 (\frac{p_3}{3})^3}_{2,N}^c
& = & 3 \cdot 2 \cdot \corr{\frac{p_2}{2}(\frac{p_3}{3})^2}_{2,N}^c \\
& = & 3 \cdot (3+3) \cdot \corr{(\frac{p_3}{3})^2}_{2,N}^c
= 18 \cdot (\frac{N}{3}+\frac{4N^3}{3}) \\
& = & 6N + 24N^3,
\een
This matches with \cite[Example 3.2.7]{Dub-Yan}.
The results of these concrete computations can also be double  checked
by the results in \cite[Appendix]{Zhou-MatMod}.

\subsection{Virasoro constraints for fat genus expansion}

If one introduces the 't Hooft coupling constant $t = Ng_s$ and
take $\tilde{T}_n = \frac{N(g_n- \delta_{n,2})}{nt}
= \frac{g_n- \delta_{n,2}}{n g_s}$,
then the Virasoro operators become:
\ben
&& L_{-1,t} = - \frac{\pd}{\pd g_1}
+ \sum_{n \geq 1} ng_{n+1} \frac{\pd}{\pd g_n} +  tg_1g_s^{-2}, \\
&& L_{0,t} = - 2\frac{\pd}{\pd g_2} + \sum_{n \geq 1} ng_n \frac{\pd}{\pd g_n}
+t^2g_s^{-2}, \\
&& L_{1,t} = - 3\frac{\pd}{\pd g_3} + \sum_{n \geq 1} (n+1)g_n \frac{\pd}{\pd g_{n+1}}
+ 2t\frac{\pd}{\pd g_1}, \\
&& L_{m,t} = \sum_{k \geq 1} (k+m) (g_k-\delta_{k,2}) \frac{\pd}{\pd g_{k+m}}
+ g_s^2 \sum_{k=1}^{m-1} k(m-k)\frac{\pd}{\pd g_k} \frac{\pd}{\pd g_{m-k}}
+ 2 tm  \frac{\pd}{\pd g_m},
\een
where $m \geq 2$.
We will refer to the corresponding constraints on the fat free energy function
as the {\em fat Virasoro constraints}.

\subsection{Fat Virasoro constraints in terms of fat correlators}
As in the thin case,
we rewrite the fat Virasoro constraints in terms of fat correllators.
The fat puncture equation can be written as
\be
\corr{p_1 \cdot \frac{p_{a_1}}{a_1} \cdots \frac{p_{a_n}}{a_n}}^c_{g}(t) =
\sum_{j=1}^n (a_j-1) \cdot \corr{\frac{p_{a_1}}{a_1} \cdots
\frac{p_{a_j-1}}{a_j-1} \cdots \frac{p_{a_n}}{a_n}}_{g}^c(t),
\ee
together with initial value:
\be
\corr{p_1^2}_{0}^c(t) = t
\ee
The fat dilaton equation can be written as
\be
\corr{p_2 \cdot \frac{p_{a_1}}{a_1} \cdots \frac{p_{a_n}}{a_n}}^c_{g}(t) =
\sum_{j=1}^n a_j \cdot \corr{\frac{p_{a_1}}{a_1} \cdots \frac{p_{a_n}}{a_n}}_{g}^c(t),
\ee
together with initial value:
\be
\corr{p_2}_{0}^c(t) = t^2.
\ee
Note in the thin case the correlator in the initial value is in genus $1$. 
The third equation in the sequence can be written  as
\be \label{eqn:Thin-Vira-3}
\begin{split}
\corr{p_3 \cdot \frac{p_{a_1}}{a_1} \cdots \frac{p_{a_n}}{a_n}}^c_{g}(t)
& =\sum_{j=1}^n (a_j+1) \cdot \corr{\frac{p_{a_1}}{a_1} \cdots
\frac{p_{a_j+1}}{a_j+1} \cdots \frac{p_{a_n}}{a_n}}_{g}^c(t) \\
& + 2t \corr{p_1 \cdot \frac{p_{a_1}}{a_1} \cdots \frac{p_{a_n}}{a_n}}^c_{g}(t).
\end{split}
\ee
Note in the second line,
the correlator is in genus $g$. 
This is different from the thin case \eqref{eqn:Thin-Vira-3}
where the corresponding correlator is in genus $g-1$.
And for $m \geq 2$,
\be
\begin{split}
& \corr{p_ {m+2} \cdot \frac{p_{a_1}}{a_1} \cdots \frac{p_{a_n}}{a_n}}^c_{g}(t)
=\sum_{j=1}^n (a_j+m) \cdot \corr{\frac{p_{a_1}}{a_1} \cdots
\frac{p_{a_j+m}}{a_j+m} \cdots \frac{p_{a_n}}{a_n}}_{g}^c(t) \\
& \qquad + 2t \corr{p_m \cdot \frac{p_{a_1}}{a_1} \cdots \frac{p_{a_n}}{a_n}}^c_{g}(t) \\
& \qquad + \sum_{k=1}^m \corr{p_kp_{m-k} \cdot \frac{p_{a_1}}{a_1} \cdots 
\frac{p_{a_n}}{a_n}}^c_{g-1}(t) \\
& \qquad + \sum_{k=1}^m \sum_{\substack{g_1+g_2=g\\I_1 \coprod I_2 = [n]}}
\corr{p_k \cdot \prod_{i\in I_1} \frac{p_{a_i}}{a_i}}^c_{g_1, N}
\cdot \corr{p_{m-k} \cdot \prod_{i\in I_2} \frac{p_{a_i}}{a_i}}^c_{g_2, N},
\end{split}
\ee
where $[n]=\{1, \dots, n\}$.

Now we present some examples of computations of fat correlators by the fat Virasoro constraints.
In degree $8$ we have:
\ben
\corr{\frac{p_3}{3}\frac{p_5}{5}}_{0}^c(t)
& = &\frac{1}{3} \cdot \biggl( 6 \corr{\frac{p_6}{6}}_{0}^c(t)
+ 2t \corr{p_1\frac{p_5}{5}}_{0}^c(t)\biggr) \\
& = & \frac{1}{3} \cdot \biggl( 6 \cdot  \frac{5t^4}{6} 
+ 2t \cdot 2t^3\biggr)  =3t^4,
\een
and in genus one,
\ben
\corr{\frac{p_3}{3}\frac{p_5}{5}}_{1}^c(t)
& = &\frac{1}{3} \cdot \biggl( 6 \corr{\frac{p_6}{6}}_{1}^c(t)
+ 2N \corr{p_1\frac{p_5}{5}}_{1}^c(t)\biggr) \\
& = & \frac{1}{3} \cdot \biggl( 6 \cdot  \frac{5t^2}{3}  
+ 2t \cdot t \biggr) 
= 4t^2.
\een
By the fat selection rule \eqref{eqn:Fat-Select},
$\corr{\frac{p_3}{3}\frac{p_5}{5}}_{g}^c(t)$
is nonvanishing only for $g=0$ and $1$.

\section{Some More Applications of the Thin Virasoro Constraints}
\label{sec:More-App}

In this Section we present some applications
of thin Virasoro constraints to compute $F_{g,N}$.

\subsection{Computation of $F_{0,N}$ by thin Virasoro constraints}

The thin Virasoro constraints in genus zero are:
\ben
&&  \frac{\pd F_{0,N}}{\pd g_1}
= \sum_{n \geq 1} ng_{n+1} \frac{\pd F_{0,N}}{\pd g_n} + N g_1, \\
&& 2\frac{\pd F_{0,N}}{\pd g_2}
= \sum_{n \geq 1} ng_n \frac{\pd F_{0,N}}{\pd g_n}, \\
&& (n+2) \frac{\pd F_{0,N}}{\pd g_{n+2}}
= \sum_{k \geq 1} (k+n) g_k \frac{\pd F_{0,N}}{\pd g_{k+n}}, \qquad n \geq 1.
\een
Together with the initial value $F_{0,N}(0, 0, \dots) = 0$,
these determine $F_{0,N}$ uniquely.
Note $f_0=\frac{1}{N} F_{0, N}$ satisfies the following recursion relations:
\bea
&&  \frac{\pd f_{0}}{\pd g_1}
= \sum_{n \geq 1} ng_{n+1} \frac{\pd f_0}{\pd g_n} + g_1, \label{eqn:f0-Vir-1} \\
&& 2\frac{\pd f_{0}}{\pd g_2}
= \sum_{n \geq 1} ng_n \frac{\pd f_0}{\pd g_n}, \label{eqn:f0-Vir0} \\
&& (n+2) \frac{\pd f_{0}}{\pd g_{n+2}}
= \sum_{k \geq 1} (k+n) g_k \frac{\pd f_0}{\pd g_{k+n}}, \qquad n \geq 1. \label{eqn:f0-Virn}
\eea
These are just the $N=1$ case of the thin Virasoro constraints in genus zero.
On the other hand,
these are exactly the Virasoro constraints for $F_0^{1D}$
of genus zero free energy of
topological 1D gravity studied in an earlier work \cite{Zhou-1D}:

\begin{thm} \cite[Theorem 7.8]{Zhou-1D}
The partition function $Z$ of topological 1D gravity satisfies the following equations for $m \geq -1$:
\be \label{eqn:Virasoro-New}
\tilde{L}_m Z = 0,
\ee
where
\bea
&& \tilde{L}_{-1} = \frac{t_0}{\lambda^2} + \sum_{m \geq 1} (t_{m}-\delta_{m,1}) \frac{\pd}{\pd t_{m-1}},
\label{eqn:Virasoro-New--1} \\
&& \tilde{L}_0 = 1 + \sum_{m \geq 0} (t_{m}-\delta_{m,1}) (m+1) \frac{\pd}{\pd t_{m}}, \\
&& \tilde{L}_1 = 2 \lambda^2 \frac{\pd}{\pd t_0}
+  \sum_{n \geq 0} (t_{n}-\delta_{n,1}) \frac{(n+2)!}{n!} \frac{\pd}{\pd t_{n+1}}, \\
&& \tilde{L}_m = 2 \lambda^2 m! \frac{\pd}{\pd t_{m-1}}
+ \lambda^4 \sum_{\substack{m_1+m_2=m\\m_1, m_2 \geq 1}}  m_1!\frac{\pd}{\pd t_{m_1-1}}
\cdot  m_2!\frac{\pd}{\pd t_{m_2-1}} \label{eqn:Virasoro-New-m} \\
&& \;\;\;\; +  \sum_{n \geq 0} (t_{n}-\delta_{n,1}) \frac{(m+n+1)!}{n!} \frac{\pd}{\pd t_{m+n}}, \nonumber
\eea
for $m \geq 2$.
Furthermore,
$\{\tilde{L}_m\}_{m \geq 1}$ satisfies the following commutation relations:
\be
[\tilde{L}_m, \tilde{L}_n ] = 0,
\ee
for $m, n \geq -1$.
\end{thm}

For $F_0^{1D}$, these Virasoro constraints give:
\bea
&& \frac{\pd F_0^{1D}}{\pd t_0}
= \sum_{m \geq 1} t_{m} \frac{\pd F_0^{1D}}{\pd t_{m-1}} + t_0, \\
&& 2\frac{\pd F_0^{1D}}{\pd t_1}
= \sum_{m \geq 0} (m+1) t_m\frac{\pd F_0^{1D}}{\pd t_{m}}, \\
&& (m+2)!\frac{\pd F_0^{1D}}{\pd t_{m+1}}
=  \sum_{n \geq 1} t_{n-1}\frac{(m+n)!}{(n-1)!}
\frac{\pd F_0^{1D}}{\pd t_{m+n-1}},
\eea
for $m \geq 1$.
After changing $t_n$ to $n! g_{n+1}$,
these match with \eqref{eqn:f0-Vir-1}-\eqref{eqn:f0-Virn}.
So we have proved the following:

\begin{thm}
The thin genus zero part $F_{0, N}$ of $F_N$ is related to $F_{0, N=1}$ in the following way:
\be \label{eqn:F0N-F0-1D}
F_{0,N} = N \cdot F_0^{1D}.
\ee
\end{thm}

Recall the following result in \cite{Zhou-1D}:

\begin{thm} \cite[Thoerem 5.6]{Zhou-1D}
The following formulas hold:
\bea
F_0^{1D} & = & \sum_{k=1}^\infty \frac{1}{k(k+1)}
\sum_{ p_1 + \cdots + p_{k+1} = k-1 } \frac{t_{p_1}}{p_1!} \cdots
\frac{t_{p_{k+1}}}{p_{k+1}!} \\
& = & \sum_{k=1}^\infty \frac{1}{k(k+1)(1-t_1)^k}
\sum_{ \substack{p_1 + \cdots + p_{k+1} = k-1 \\ p_1, \dots, p_{k+1} \neq 1}}
\frac{t_{p_1}}{p_1!} \cdots
\frac{t_{p_{k+1}}}{p_{k+1}!}.
\eea
\end{thm}

 And so as a consequence,
 we get:

\begin{thm}
The following formulas hold:
\bea
F_{0,N} & = & N \sum_{k=1}^\infty \frac{1}{k(k+1)}
\sum_{ \substack{j_1 + \cdots + j_{k+1} = 2k \\j_1, \dots, j_{k+1} \geq 1} }
g_{j_1}\cdots g_{j_{k+1}} \\
& = & N  \sum_{k=1}^\infty \frac{1}{k(k+1)(1-g_2)^k}
\sum_{ \substack{j_1 + \cdots + j_{k+1} = 2k \\ j_1, \dots, j_{k+1} \neq 2}}
g_{j_1} \cdots g_{j_{k+1}}.
\eea
\end{thm}

For example,
\ben
F_{0,N} & = & \frac{N}{2} g_1^2
+ \frac{N}{2} g_2g_1^2
+ \frac{N}{3} g_3g_1^3 + \frac{N}{2} g_2^2g_1^2 + \cdots \\
& = & \frac{N}{2}\frac{g_1^2}{1-g_2} + \frac{N}{3} \frac{g_1^3g_3}{(1-g_2)^2}
+ \frac{N}{4} \frac{g_1^4g_4}{(1-g_2)^3}
+ \frac{N}{5} \frac{g_1^5g_5}{(1-g_2)^4} \\
& + &  \frac{N}{2}\frac{g_1^4g_2^2}{(1-g_2)^4}
+ \cdots,
\een
these match with the first few terms of $F_{0, N}$ given in \S \ref{sec:Thin}.

\subsection{Computations for $F_{1,N}$ by thin Virasoro constraints}

In the above we have computed $F_{0, N}$ by Virasoro constraints and
identified it with $NF^{1D}$.
Now apply the same idea to compute $F_{1,N}$.
The Virasoro constraints for $F_{1, N}$ are:
\ben
&& \frac{\pd F_{1,N}}{\pd g_1}
= \sum_{n \geq 1} ng_{n+1} \frac{\pd F_{1,N}}{\pd g_n},\\
&& 2\frac{\pd F_{1,N}}{\pd g_2} = \sum_{n \geq 1} ng_n \frac{\pd F_{1,N}}{\pd g_n} + N^2, \\
&& 3\frac{\pd F_{1,N}}{\pd g_2} = \sum_{n \geq 1} ng_n \frac{\pd F_{1,N}}{\pd g_{n+1}}
+ 4N \frac{\pd F_{0,N}}{\pd g_1}, \\
&& (n+2) \frac{\pd F_{1,N}}{\pd g_{n+2}}
= \sum_{k \geq 1} (k+n) g_k \frac{\pd F_{1,N}}{\pd g_{k+n}}
+ \sum_{k=1}^{n} k(n-k)\frac{\pd F_{0,N}}{\pd g_k} \frac{\pd  F_{0,N}}{\pd g_{n-k}}
+ 2Nn \frac{\pd F_{0,N}}{\pd g_n},
\een
where $n\geq 2$.
By taking $N=1$,
we get the Virasoro constraints of $F_1^{1D}$, because we have $F_{g, N=1}= F^{1D}_g$.
It follows that if one sets $f_1 = N^{-2}F_{1,N}$ and $f_0=N^{-1}F_{0,N}$,
then dividing by $N^2$ on both sides of the above equalities,
one sees that $f_0$ and $f_1$ satisfy the Virasoro constraints and initial value conditions
for $F^{1D}_1$ and $F^{1D}_0$ respectively.
Therefore, we have
\be \label{eqn:F1N-F1-1D}
F_{1,N} = N^2 \cdot F^{1D}_1.
\ee

\subsection{Higher genera case}
However,
we do not have
\ben
F_{g, N} = N^{g+1}\cdot F^{1D}_g
\een
for $g \geq 2$.
For example,
the Virasoro constraints for $F_{2,N}$ for $n=-1,0,1$ are
\ben
&& \frac{\pd F_{2,N}}{\pd g_1}
= \sum_{n \geq 1} ng_{n+1} \frac{\pd F_{2,N}}{\pd g_n},\\
&& 2\frac{\pd F_{2,N}}{\pd g_2} = \sum_{n \geq 1} ng_n \frac{\pd F_{2,N}}{\pd g_n}, \\
&& 3\frac{\pd F_{2,N}}{\pd g_3} = \sum_{n \geq 1} ng_n \frac{\pd F_{2,N}}{\pd g_{n+1}}
+ 4N \frac{\pd F_{1,N}}{\pd g_1}.
\een
These are similar to the case of $F_{1,N}$.
But for $n \geq 2$,
\ben
&& (n+2) \frac{\pd F_{2,N}}{\pd g_{n+2}}
= \sum_{k \geq 1} (k+n) g_k \frac{\pd F_{2,N}}{\pd g_{k+n}}
+ 2 \sum_{k=1}^{n} k(n-k)\frac{\pd F_{0,N}}{\pd g_k} \frac{\pd  F_{1,N}}{\pd g_{n-k}}
+ 2Nn \frac{\pd F_{1,N}}{\pd g_n} \\
&& \qquad \qquad  \;\;\; + \sum_{k=1}^{n} k(n-k)\frac{\pd^2 F_{0,N}}{\pd g_k \pd g_{n-k}},
\een
the extra term on the second line spoils the homogeneity in $N$.
Write
\be
F_{2, N} = f_{2,1}N +f_{2,3}N^3,
\ee
then one gets for $f_{2,3}$ the following recursion relations:
\ben
&& \frac{\pd f_{2,3}}{\pd g_1}
= \sum_{n \geq 1} ng_{n+1} \frac{\pd f_{2,3}}{\pd g_n},\\
&& 2\frac{\pd f_{2,3}}{\pd g_2} = \sum_{n \geq 1} ng_n \frac{\pd f_{2,3}}{\pd g_n}, \\
&& 3\frac{\pd f_{2,3}}{\pd g_3} = \sum_{n \geq 1} ng_n \frac{\pd f_{2,3}}{\pd g_{n+1}}
+ 4 \frac{\pd f_{1}}{\pd g_1},
\een
and for $n \geq 2$,
\ben
&& (n+2) \frac{\pd f_{2,3}}{\pd g_{n+2}}
= \sum_{k \geq 1} (k+n) g_k \frac{\pd f_{2,3}}{\pd g_{k+n}}
+ 2 \sum_{k=1}^{n} k(n-k)\frac{\pd f_{0}}{\pd g_k} \frac{\pd  f_{1}}{\pd g_{n-k}}
+ 2n \frac{\pd f_{1}}{\pd g_n},
\een
and for $f_{2,1}$ one gets the following recursion relations:
\ben
&& \frac{\pd f_{2,1}}{\pd g_1}
= \sum_{n \geq 1} ng_{n+1} \frac{\pd f_{2,1}}{\pd g_n},\\
&& 2\frac{\pd f_{2,1}}{\pd g_2} = \sum_{n \geq 1} ng_n \frac{\pd f_{2,1}}{\pd g_n}, \\
&& 3\frac{\pd f_{2,N}}{\pd g_3} = \sum_{n \geq 1} ng_n \frac{\pd f_{2,1}}{\pd g_{n+1}},
\een
and for $n \geq 2$,
\ben
&& (n+2) \frac{\pd f_{2,1}}{\pd g_{n+2}}
= \sum_{k \geq 1} (k+n) g_k \frac{\pd f_{2,1}}{\pd g_{k+n}}
+ \sum_{k=1}^{n} k(n-k)\frac{\pd^2 f_{0\emph{}}}{\pd g_k \pd g_{n-k}}.
\een
One can use these recursion relations to compute $f_{2,3}$ and $f_{2,1}$
from $f_0 = F_0^{1D}$ and $f_1 = F_1^{1D}$.

\subsection{An application of the dilaton equation}

Now we generalize the result of \cite[\S 6.2]{Zhou-1D} to matrix models.
The dilaton equation
\be \label{eqn:Dilaton}
L_{0,N}Z_N = 0
\ee
can be rewritten as
\be
\frac{\pd F_N}{\pd g_2} = \sum_{m \geq 1} \frac{m}{2} g_m \frac{\pd F}{\pd g_m} + \frac{N^2}{2}.
\ee
In terms of correlators,
\bea
&& \corr{\frac{p_2}{2}}_{1,N}^c = \frac{N^2}{2}, \\
&& \corr{\frac{p_2}{2} \prod_{j=1}^n \frac{p_{a_j}}{a_j}}_{g,N}^c
= \sum_{j=1}^n \frac{a_j}{2} \corr{\prod_{j=1}^n \frac{p_{a_j}}{a_j}}^c_{g,N}.
\eea
Therefore,
\be
\corr{(\frac{p_2}{2})^m}_{1,N}^c = \frac{N^2}{2} (m-1)!,
\ee
and for $a_2, \dots, a_n \neq 1$ which satisfies the selection rule \eqref{eqn:Thin-Select},
$$a_1+ \cdots + a_n = 2g-2+2n,$$
we have
\be
\corr{(\frac{p_2}{2})^m \prod_{j=1}^n \frac{p_{a_j}}{a_j} }_{g,N}^c
= \prod_{k=0}^{m-1} (g-1+n +k) \cdot \corr{\prod_{j=1}^n \frac{p_{a_j}}{a_j} }_{g,N}^c.
\ee
It follows that we have

\begin{thm} \label{thm:1-t1}
The free energy $F_N$  can be rewritten in the following form:
\be
F_N = \half \log (1-g_2)
+ \sum_{g\geq 0, n> 0} \sum_{\substack{a_1, \dots, a_n \neq 2
\\ a_1+ \cdots + a_n = 2g-2+2n }}
\frac{\corr{\prod_{j=1}^n \frac{p_{a_j}}{a_j} }_N^c}{(1-t_2)^{g-1+n}}
\frac{1}{n!} \prod_{j=1}^n g_{a_j}.
\ee
\end{thm}

For example,
\ben
F_{0,N} & = & \sum_{n> 0} \sum_{\substack{a_1, \dots, a_n \neq 2
\\ a_1+ \cdots + a_n = 2n-2 }}
\frac{\corr{\prod_{j=1}^n \frac{p_{a_j}}{a_j} }_{0,N}^c}{(1-t_2)^{n-1}}
\frac{1}{n!} \prod_{j=1}^n g_{a_j}, \\
F_{1, N} & = & \log (1-g_2) +\sum_{n> 0} \sum_{\substack{a_1, \dots, a_n \neq 2
\\ a_1+ \cdots + a_n = 2n }}
\frac{\corr{\prod_{j=1}^n \frac{p_{a_j}}{a_j} }_{1,N}^c}{(1-t_2)^{n}}
\frac{1}{n!} \prod_{j=1}^n g_{a_j}, \\
F_{g, N} & = &\sum_{n> 0} \sum_{\substack{a_1, \dots, a_n \neq 2
\\ a_1+ \cdots + a_n = 2g-2+2n }}
\frac{\corr{\prod_{j=1}^n \frac{p_{a_j}}{a_j} }_{g,N}^c}{(1-t_2)^{g-1+n}}
\frac{1}{n!} \prod_{j=1}^n g_{a_j}, \qquad g >1.
\een
In particular,
by the dilaton equation \eqref{eqn:Dilaton},
one can reduce the calculations of $F_{g, N}$ to the calculations of correlators
$\corr{p_{a_1} \cdots p_{a_n}}_N$ with $a_i \neq 2$, $i =1, \dots n$.
Similarly,
one can use the string equation
\be \label{eqn:String}
L_{-1, N} Z_N = 0
\ee
to further reduce to the correlators
$\corr{p_{a_1} \cdots p_{a_n}}_N$ with $a_i > 2$, $i =1, \dots n$.
An even better way to use such ideas is to introduce suitable changes of
coordinates on the space of coupling constants to be discussed
in next Section.

\section{Thin Virasoro Constraints in Renormalized Coupling Constants}
\label{sec:Renormalized}

We now combine the Virasoro constraints with the renormalized coupling constants
to prove some structure results for $F_{g,N}$.

\subsection{The renormalized coupling constants}

Let us recall the following coordinate change on the big phase space of coupling constants:
\bea
&& I_0 = \sum_{k=1}^\infty \frac{1}{k}
\sum_{p_1 + \cdots + p_k = k-1} \frac{t_{p_1}}{p_1!} \cdots
\frac{t_{p_k}}{p_k!}, \label{eqn:I0} \\
&& I_k= \sum_{n \geq 0} t_{n+k} \frac{I_0^n}{n!}, \;\;\;\; k \geq 1. \label{eqn:Ik}
\eea
These will be referred to as the {\em renormalized coupling constants}.
These series were introduced in \cite{Itzykson-Zuber}
to express the free energy of topological 2D gravity.
In \cite{Zhou-1D}
they were  understood  as new coordinates on the big phase space.
By \cite[Proposition 2.4]{Zhou-1D},
\be \label{eqn:T-in-I}
t_k = \sum_{n=0}^\infty \frac{(-1)^n I_0^n}{n!}I_{n+k}.
\ee
The renormalized coupling constants
were used to gain better understanding of
the global nature of the behavior of the theory
on the big phase space.
For example, two different methods were used to show that for topological 1D gravity,
\bea
&& F_0^{1D} = \sum_{k=0}^\infty  \frac{(-1)^k}{(k+1)!} (I_k+\delta_{k,1}) I_0^{k+1},
\label{eqn:F0-1D}\\
&& F_1^{1D} = \frac{1}{2} \ln \frac{1}{1- I_1}, \label{eqn:F1-1D}\\
&& F_g^{1D}
  =  \sum_{\sum\limits_{j=2}^{2g-1}  \frac{j-1}{2} l_j = g-1}
 \corr{\tau_2^{l_2} \cdots \tau_{2g-1}^{l_{2g-1}}}_g
\prod_{j=2}^{2g-1} \frac{1}{l_j!}\biggl( \frac{I_j}{(1-I_1)^{(j+1)/2}}\biggr)^{l_j}, \;\; g \geq 2.
\eea

\subsection{Puncture operator $L_{-1,N}$  in I-coordinates}
As an application of \eqref{eqn:F0N-F0-1D} and \eqref{eqn:F0-1D},
we have the following result:
\be
F_{0,N} = N \sum_{k=0}^\infty  \frac{(-1)^k}{(k+1)!} (I_k+\delta_{k,1}) I_0^{k+1}.
\ee
We now give a direct proof of it in the spirit of \cite{Zhou-1D}.
By \cite[(34)]{Zhou-1D} and \eqref{eqn:T-in-I},
we have
\be \label{eqn:L-1-in-I}
L_{-1,N} = - \frac{\pd}{\pd I_0} + \frac{N}{g_s} \sum_{n=0}^\infty \frac{(-1)^nI_0^n}{n!}I_n.
\ee

It follows from \eqref{eqn:L-1-in-I} that
\bea
&& \frac{\pd F_{0,N}}{\pd I_0} = N\sum_{n=0}^\infty \frac{(-1)^nI_0^n}{n!}I_n, \\
&& \frac{\pd F_{g,N}}{\pd I_0} = 0, \;\;\; g \geq 1.
\eea
Therefore,
we get the following results generalizing \cite[Theorem  6.4]{Zhou-1D}:

\begin{thm} \label{thm:F0}
The thin genus zero part $F_{0,N}$ of the free energy $F_N$  is given in I-coordinates by:
\be
F_{0,N}= N(\frac{1}{2} I_0^2 + \sum_{n=0}^\infty \frac{(-1)^nI_0^{n+1}}{(n+1)!}I_n).
\ee
Furthermore, when $g \geq 1$,
$F_{g,N}$ is independent of $I_0$.
\end{thm}

\subsection{Dilaton operator $L_{0,N}$ in I-coordinates}

Similar to \cite[Lemma 6.5]{Zhou-1D},
the dilaton operator $L_{0,N}$ is given in I-coordinates by:
\be
L_{0,N} = - I_0 \frac{\pd}{\pd I_0} - 2 \frac{\pd}{\pd I_1}
+ \sum_{l \geq 1} (l+1) I_l \frac{\pd}{\pd I_l} + N^2.
\ee

From the dilaton equation,
one gets
\bea
&&  \frac{\pd F_{0,N}}{\pd I_1}
= \sum_{l \geq 1} \frac{l+1}{2} I_l \frac{\pd F_{0,N}}{\pd I_l}
-  \frac{1}{2} \frac{\pd F_{0,N}}{\pd I_0}, \label{eqn:Dilaton-0} \\
&&  \frac{\pd F_{1,N}}{\pd I_1}
= \sum_{l \geq 1} \frac{l+1}{2} I_l \frac{\pd F_{1,N}}{\pd I_l}
+  \frac{N^2}{2},
\label{eqn:Dilaton-1} \\
&& \frac{\pd F_{g,N}}{\pd I_1}
= \sum_{l \geq 1} \frac{l+1}{2} I_l \frac{\pd F_{g,N}}{\pd I_l}, \;\;\; g \geq 2.
\label{eqn:Dilaton-g}
\eea
By slightly modifying the proof of
\cite[Theorem 6.6]{Zhou-1D},
one can prove the following:

\begin{thm} \label{thm:Fgeq1}
In $I$-coordinates we have
\be
F_{1,N} = \frac{N^2}{2} \ln \frac{1}{1- I_1}
\ee
and for $g \geq 1$,
\be \label{eqn:FgN}
F_{g,N}  =  \sum_{\sum_{j=3}^{2g}  (j-2)l_j = 2g-2}
 \corr{p_3^{l_3} \cdots p_{2g}^{l_{2g}}}^c_{g,N}
\prod_{j=3}^{2g} \frac{1}{l_j!(j-1)!^{l_j}}\biggl( \frac{I_{j-1}}{(1-I_1)^{j/2}}\biggr)^{l_j}.
\ee
\end{thm}

By looking at the summation in \eqref{eqn:FgN},
one sees surprisingly that it is a summation over partition over $2g-2$.
Introducing new variables $q_n$ that plays the role of Newton power function $p_n$:
\be \label{eqn:qn}
q_n : = \frac{I_{n+1}}{(1-I_1)^{(n+2)/2}},
\ee
one can rewrite \eqref{eqn:FgN} as follows:
\be
F_{g,N}  =  \sum_{\sum_{k=1}^{2g-2}  km_k = 2g-2}
 \corr{p_3^{m_1} \cdots p_{2g}^{m_{2g-2}}}^c_{g,N}
\prod_{k=1}^{2g-2} \frac{1}{m_k!(k+1)!^{m_k}}q_k^{m_k}.
\ee
For example,
for $g=2$,
\ben
F_{2,N} & = & \frac{1}{2!} \corr{(\frac{p_3}{3})^2}^c_{2,N} \frac{I_2^2}{(2!)^2(1-I_1)^3}
+ \corr{\frac{p_4}{4}}_{2,N}^c \frac{I_3}{3!(1-I_1)^2} \\
& = & (\frac{2N^3}{3}+\frac{N}{6}) \frac{I_2^2}{2!^2(1-I_1)^3}
+ (\frac{1}{2}N^3 +\frac{1}{4} N) \frac{I_3}{3!(1-I_1)^2}.
\een
In terms of the new variables $q_n$:
\ben
F_{2,N}& = & (\frac{N^3}{6}+ \frac{N}{24})q_1^2
+ (\frac{N^3}{12}+\frac{N}{24}) q_2
\een
Similarly, for $g=3$,
\ben
F_{3,N} & = & \frac{1}{4!}\corr{(\frac{p_3}{3})^4}_{3,N}^c
\cdot \frac{I_2^4}{(2!)^4(1-I_1)^6}
+ \frac{1}{2!}\corr{(\frac{p_3}{3})^2\frac{p_4}{4} }_{3,N}^c \cdot \frac{I_2^2I_3}{(2!)^23!(1-I_1)^5} \\
&& + \frac{1}{2!} \corr{\frac{p_4}{4})^2}_{3,N}^c \cdot \frac{I_3^2}{(3!)^2(1-I_1)^4}
 + \corr{\frac{p_3}{3}\frac{p_5}{5} }_{3,N}^c \cdot \frac{I_2I_4}{2!4!(1-I_1)^4} \\
&&  + \corr{\frac{p_6}{6}}_{3,N}^c \frac{I_5}{5!(1-I_1)^3}.
\een
By computing the correlators, one gets:
\ben
F_{3,N} & = & \frac{1}{4!}(56N^2+64N^4)
\cdot \frac{I_2^4}{(2!)^4(1-I_1)^6}
+ \frac{1}{2!}(13N^2+12N^4) \cdot \frac{I_2^2I_3}{(2!)^23!(1-I_1)^5} \\
&& + \frac{1}{2!} \cdot \frac{1}{4}(15N^2 +9 N^4)\cdot \frac{I_3^2}{(3!)^2(1-I_1)^4}
 + (4N^2+3N^4) \cdot \frac{I_2I_4}{2!4!(1-I_1)^4} \\
&&  + (\frac{5N^2}{3}+\frac{5N^4}{6}) \frac{I_5}{5!(1-I_1)^3}.
\een
In the $q_n$ variables:
\ben
F_{3,N} & = & (\frac{1}{6}N^4+\frac{7}{48}N^2) q_1^4
+(\frac{1}{4}N^4+\frac{13}{48}N^2) q_1^2q_2
+ (\frac{1}{32}N^4+\frac{5}{96}N^2) q_2^2 \\
& + & (\frac{1}{16}N^4+\frac{1}{12}N^2) q_1q_3
+ + (\frac{1}{144}N^4+\frac{1}{72}N^2) q_4.
\een
For $g=4$,
\ben
F_{4,N} & = & \corr{(\frac{p_3}{3})^6}_{4,N}^c \cdot \frac{1}{6!}\frac{I_2^6}{(2!)^6(1-I_1)^9}
+ \corr{(\frac{p_3}{3})^4\frac{p_4}{4} }_{4,N}^c
\cdot \frac{1}{4!}\frac{I_2^4I_3}{(2!)^43!(1-I_1)^8} \\
& + &  \corr{(\frac{p_3}{3})^2(\frac{p_4}{4})^2}_{4,N}^c
\cdot \frac{1}{2!2!}\frac{I_2^2I_3^2}{(2!)^2(3!)^2(1-I_1)^7}
+ \corr{(\frac{p_4}{4})^3}_{4,N}^c \frac{1}{3!} \frac{I_3^3}{(3!)^3(1-I_1)^6} \\
& + & \corr{(\frac{p_3}{3})^3\frac{p_5}{5}}_{4,N}^c
\cdot \frac{1}{3!}\frac{I_2^3I_4}{(2!)^34!(1-I_1)^7}
+ \corr{\frac{p_3}{3}\frac{p_4}{4}\frac{p_5}{5}}_{4,N}^c \cdot \frac{I_2I_3I_4}{2!3!4!(1-I_1)^6}\\
& + & \corr{(\frac{p_5}{5})^2}_{4,N}^c \cdot \frac{1}{2!} \frac{I_4^2}{(4!)^2(1-I_1)^5}
+ \corr{(\frac{p_3}{3})^2\frac{p_6}{6}}_{4,N}^c \cdot \frac{1}{2!} \frac{I_2^2I_5}{(2!)^25!(1-I_1)^6} \\
& + & \corr{\frac{p_3}{3}\frac{p_7}{7}}_{4,N}^c \cdot \frac{I_2I_6}{2!6!(1-I_1)^5}
+ \corr{\frac{p_8}{8}}_{4,N}^c \cdot \frac{I_7}{7!(1-I_1)^4}.
\een
By computing the correlators, one gets:
\ben
F_{4,N} & = & (4736N^5+7104N^3+840N) \cdot \frac{1}{6!}\frac{I_2^6}{(2!)^6(1-I_1)^9} \\
& + & (1632N^5+3648N^3+630N)
\cdot \frac{1}{4!}\frac{I_2^4I_3}{(2!)^43!(1-I_1)^8} \\
& + &  (156N^5+384N^3+70N)
\cdot \frac{1}{2!2!}\frac{I_2^2I_3^2}{(2!)^2(3!)^2(1-I_1)^7} \\
& + & (27N^5+99N^3+\frac{45}{2}N) \frac{1}{3!} \frac{I_3^3}{(3!)^3(1-I_1)^6} \\
& + & (240N^5+612N^3+114N)
\cdot \frac{1}{3!}\frac{I_2^3I_4}{(2!)^34!(1-I_1)^7} \\
& + & (18N^5+54N^3+12N) \cdot \frac{I_2I_3I_4}{2!3!4!(1-I_1)^6}\\
& + & (\frac{36}{5}N^5+24N^3+\frac{33}{5}N) \cdot \frac{1}{2!} \frac{I_4^2}{(4!)^2(1-I_1)^5} \\
& + & (25N+\frac{370}{3}N^3 + 40N^5) \cdot \frac{1}{2!} \frac{I_2^2I_5}{(2!)^25!(1-I_1)^6} \\
& + & (7N+30N^3+8N^5) \cdot \frac{I_2I_6}{2!6!(1-I_1)^5}\\
& + & \frac{1}{8}(21N+70N^3+14N^5) \cdot \frac{I_7}{7!(1-I_1)^4}.
\een
In the $q_n$-coordinates,
\ben
F_{4,N} & =& (\frac{37}{360}N^5+\frac{37}{240}N^3+\frac{7}{384}N)q_1^6
+ (\frac{17}{24}N^5+\frac{19}{12}N^3+\frac{35}{128}N)q_1^4q_2 \\
& + & (\frac{13}{48}N^5+\frac{2}{3}N^3+\frac{35}{288}N) q_1^2q_2^2
+ (\frac{1}{48}N^5+\frac{11}{144}N^3+\frac{5}{288}N) q_2^3\\
&+& ((\frac{5}{24}N^5+\frac{17}{32}N^3+\frac{19}{192}N)q_1^3q_3
+ (\frac{1}{16}N^5+\frac{3}{16}N^3+\frac{1}{24}N)q_1q_2q_3 \\
&+& (\frac{1}{160}N^5+\frac{1}{48}N^3+\frac{11}{1920}N) q_3^2
+ (\frac{5}{192}N^5+\frac{37}{288}N^3+\frac{1}{24}N )q_1^2q_4 \\
&+& (\frac{7}{1440} N^5+\frac{1}{48} N^3+\frac{1}{180} N) q_1q_5
+ (\frac{1}{1920}N^5+\frac{1}{576}N^3+\frac{1}{2880}N) q_6.
\een

\section{Renoramlizations of Hermitian One-Matrix Models}
\label{sec:Renormalization}

We first recall the renormalization of the universal action function
studied in \cite{Zhou-1D},
then apply it to Hermitian one-matrix models to give another proof
of the mains results in last Section.

\subsection{Renormalization of the universal action function}

By the {\em universal action function}
we mean the following formal power series in $x$ depending on infinitely
many parameters $t_0, \dots, t_n, \dots$:
\be \label{eqn:Action}
S = - \frac{1}{2}x^2 + \sum_{n \geq 1} t_{n-1} \frac{x^n}{n!},
\ee
The coefficients $t_n$'s will be called the {\em coupling constants}.
Let $x_\infty$ be the solution of
$$\frac{\pd S}{\pd x}  = 0,$$
i.e., $x_\infty$ satisfies the   equation:
\be \label{eqn:Critical}
x_\infty = \sum_{n\geq 0} t_n \frac{x_\infty^n}{n!}.
\ee
By \cite[Proposition 2.2]{Zhou-1D},
the following formula for $x_\infty$ holds:
\be \label{eqn:Xinfinity}
x_\infty = \sum_{k=1}^\infty \frac{1}{k}
\sum_{p_1 + \cdots + p_k = k-1} \frac{t_{p_1}}{p_1!} \cdots
\frac{t_{p_k}}{p_k!}.
\ee
By \cite[Theorem 2.3, (56)]{Zhou-1D},
\be
S(x) =\sum_{k=0}^\infty  \frac{(-1)^k}{(k+1)!} (I_k+\delta_{k,1}) I_0^{k+1}
+ \sum_{n=2}^\infty \frac{I_{n-1}-\delta_{n,2} }{n!} (x-I_0)^n,
\ee
where $I_0 = x_\infty$ and $I_k$ ($k\geq 1$) have already been
recalled earlier in \eqref{eqn:I0} and \eqref{eqn:Ik}.

\subsection{Renormalization of Hermitian one-matrix models}

Now in \eqref{eqn:Log-gas} we take
$\tilde{T}_n = \frac{t_{n-1}-\delta_{n,2}}{n!g_s}$ to get
\ben
Z_N &= & c_N\int_{\bR^N} \prod^N_{i=1}
d\lambda_i \cdot \exp \biggl( \sum_{i=1}^N
\sum_{n=1}^\infty \frac{t_{n-1}-\delta_{n,2}}{n!g_s} \lambda^n_i \biggr)
\cdot \prod_{1\leq i<j \leq N} (\lambda_i - \lambda_j)^2 \\
& = &  c_N\int_{\bR^N} \prod^N_{i=1}
d\lambda_i \cdot \exp \sum_{i=1}^N \frac{1}{g_s} \biggl(
\sum_{k=0}^\infty  \frac{(-1)^k}{(k+1)!} (I_k+\delta_{k,1}) I_0^{k+1} \\
&& + \sum_{n=2}^\infty \frac{I_{n-1}-\delta_{n,2} }{n!} (\lambda_i-I_0)^n \biggr)
\cdot \prod_{1\leq i<j \leq N} (\lambda_i - \lambda_j)^2 \\
& = & \exp \sum_{i=1}^N \frac{1}{g_s}\biggl(
\sum_{k=0}^\infty  \frac{(-1)^k}{(k+1)!} (I_k+\delta_{k,1}) I_0^{k+1} \biggr) \\
&& \cdot c_N\int_{\bR^N} \prod^N_{i=1}
d\lambda_i  \cdot \exp \sum_{i=1}^N \frac{1}{g_s}
\biggl( \sum_{n=2}^\infty \frac{I_{n-1}-\delta_{n,2} }{n!} \lambda_i^n \biggr)
\cdot \prod_{1\leq i<j \leq N} (\lambda_i - \lambda_j)^2 \\
& = & \exp \biggl(\frac{N}{g_s}
\sum_{k=0}^\infty  \frac{(-1)^k}{(k+1)!} (I_k+\delta_{k,1}) I_0^{k+1} \biggr)
\cdot (1-I_1)^{-N^2/2} \\
&& \cdot c_N\int_{\bR^N} \prod^N_{i=1}
d\lambda_i  \cdot \exp \sum_{i=1}^N \frac{1}{g_s}
\biggl(-\frac{1}{2}\lambda_i^2
+ \sum_{n=3}^\infty \frac{I_{n-1}}{n!(1-I_1)^{n/2}} \lambda_i^n \biggr)
\cdot \prod_{1\leq i<j \leq N} (\lambda_i - \lambda_j)^2 \\
& = & \exp \biggl(\frac{N}{g_s}
\sum_{k=0}^\infty  \frac{(-1)^k}{(k+1)!} (I_k+\delta_{k,1}) I_0^{k+1} \biggr)
\cdot (1-I_1)^{-N^2/2} \\
&& \cdot \frac{\int_{\cH_N}dM
\cdot \exp \sum_{i=1}^N \frac{1}{g_s}
\biggl(-\frac{1}{2}\tr (M^2)
+ \sum_{n=3}^\infty \frac{I_{n-1}}{n!(1-I_1)^{n/2}} \tr(M^n) \biggr) }
{\int_{\cH_N} dM \exp\biggl(-\frac{1}{2g_s} \tr (M^2)\biggr)}.
\een
So we have proved the following generalization of
\cite[(124)]{Zhou-1D}:

\begin{thm}
The partition function $Z_N$ of Hermitian $N\times$-matrix models
remains the same under the following renormalizations of coupling constants:
\be
\begin{split}
Z_N & = \exp \biggl(\frac{N}{g_s}
\sum_{k=0}^\infty  \frac{(-1)^k}{(k+1)!} (I_k+\delta_{k,1}) I_0^{k+1}
+ \frac{N^2}{2} \log \frac{1}{1-I_1} \biggr) \\
& \cdot Z_N|_{g_1=g_2 =0, g_n = \frac{I_{n-1}}{(n-1)!(1-I_1)^{n/2}}, n \geq 3}.
\end{split}
\ee
\end{thm}

From this Theorem,
one can rederive Theorem \ref{thm:F0} and Theorem \ref{thm:Fgeq1}.

\section{From Thin Genus Expansion to Fat Genus Expansion}
\label{sec:Fat}

In this Section we show how to get the fat genus expansion from the thin genus expansion.

Recall the thin genus expansion is of the form:
\be
F_N = g_s^{-1} F_{0, N} + F_{1, N}
 + g_s F_{2, N} + g_s^2 F_{3, N} + g_s^3 F_{4, N} + \cdots,
\ee
where each $F_{g,N}$ is a polynomial in $N$.
Now we change $N$ to $tg_s^{-1}$,
where $t= Ng_s$ is the 't Hooft coupling constant:
\ben
&& \sum_{g \geq 0} g_s^{2g-2} F_g(t) \\
& = & g_s^{-2} t (\frac{1}{2} I_0^2
+ \sum_{n=0}^\infty \frac{(-1)^nI_0^{n+1}}{(n+1)!}I_n)
+ t^2 \log \frac{1}{1-I_1} \\
& + & g_s\biggl((\frac{N^3}{6}+ \frac{N}{24})q_1^2
+ (\frac{N^3}{12}+\frac{N}{24}) q_2 \biggr) \\
& + & g_s^2\biggl(\frac{1}{4!}(56t^2g_s^{-2}+64t^4g_s^{-4})
\cdot \frac{I_2^4}{(2!)^4(1-I_1)^6}
+ \frac{1}{2!}(13t^2g_s^{-2}+12t^4g_s^{-4}) \cdot \frac{I_2^2I_3}{(2!)^23!(1-I_1)^5} \\
&& + \frac{1}{2!} \cdot \frac{1}{4}(15t^2g_s^{-2} +9 t^4g_s^{-4})\cdot \frac{I_3^2}{(3!)^2(1-I_1)^4}
 + (4t^2g_s^{-2}+3t^4g_s^{-4}) \cdot \frac{I_2I_4}{2!4!(1-I_1)^4} \\
&&  + (\frac{5t^2g_s^{-2}}{3}+\frac{5t^4g_s^{-4}}{6}) \frac{I_5}{5!(1-I_1)^3} \biggr)
\een
plus the contributions from $F_{2,N}$
\ben
& + & g_s^3 \biggl[ (4736t^5g_s^{-5}+7104t^3g_s^{-3}+840tg_s^{-1}) \cdot \frac{1}{6!}\frac{I_2^6}{(2!)^6(1-I_1)^9} \\
& + & (1632t^5g_s^{-5}+3648t^3g_s^{-3}+630tg_s^{-1})
\cdot \frac{1}{4!}\frac{I_2^4I_3}{(2!)^43!(1-I_1)^8} \\
& + &  (156t^5g_s^{-5}+384t^3g_s^{-3}+70tg_s^{-1})
\cdot \frac{1}{2!2!}\frac{I_2^2I_3^2}{(2!)^2(3!)^2(1-I_1)^7} \\
& + & (27t^5g_s^{-5}+99t^3g_s^{-3}+\frac{45}{2}tg_s^{-1}) \frac{1}{3!} \frac{I_3^3}{(3!)^3(1-I_1)^6} \\
& + & (240t^5g_s^{-5}+612t^3g_s^{-3}+114tg_s^{-1})
\cdot \frac{1}{3!}\frac{I_2^3I_4}{(2!)^34!(1-I_1)^7} \\
& + & (18t^5g_s^{-5}+54t^3g_s^{-3}+12tg_s^{-1}) \cdot \frac{I_2I_3I_4}{2!3!4!(1-I_1)^6}\\
& + & (\frac{36}{5}t^5g_s^{-5}+24t^3g_s^{-3}+\frac{33}{5}tg_s^{-1}) \cdot \frac{1}{2!} \frac{I_4^2}{(4!)^2(1-I_1)^5} \\
& + & (25tg_s^{-1}+\frac{370}{3}t^3g_s^{-3} + 40t^5g_s^{-5}) \cdot \frac{1}{2!} \frac{I_2^2I_5}{(2!)^25!(1-I_1)^6} \\
& + & (7tg_s^{-1}+30t^3g_s^{-3}+8t^5g_s^{-5}) \cdot \frac{I_2I_6}{2!6!(1-I_1)^5}\\
& + & \frac{1}{8}(21tg_s^{-1}+70t^3g_s^{-3}+14t^5g_s^{-5}) \cdot \frac{I_7}{7!(1-I_1)^4}
\biggr] +\cdots.
\een
Therefore we get:
\ben
F_0(t)
& = & t (\frac{1}{2} I_0^2
+ \sum_{n=0}^\infty \frac{(-1)^nI_0^{n+1}}{(n+1)!}I_n) \\
& + & t^3\biggl(\frac{2}{3} \frac{I_2^2}{2!^2(1-I_1)^3}
+ \frac{1}{2} \frac{I_3}{3!(1-I_1)^2} \biggr) \\
& + & t^4\biggl(\frac{1}{4!} \cdot 64
\cdot \frac{I_2^4}{(2!)^4(1-I_1)^6}
+ \frac{1}{2!}\cdot 12  \cdot \frac{I_2^2I_3}{(2!)^23!(1-I_1)^5} \\
&& + \frac{1}{2!} \cdot \frac{1}{4} \cdot 9  \cdot \frac{I_3^2}{(3!)^2(1-I_1)^4}
 +  3  \cdot \frac{I_2I_4}{2!4!(1-I_1)^4}
 +  \frac{5}{6} \cdot \frac{I_5}{5!(1-I_1)^3} \biggr)
 \een
plus
\ben
& + & t^5 \biggl[ 4736 \cdot \frac{1}{6!}\frac{I_2^6}{(2!)^6(1-I_1)^9}
+  1632 \cdot \frac{1}{4!}\frac{I_2^4I_3}{(2!)^43!(1-I_1)^8} \\
& + &  156 \cdot \frac{1}{2!2!}\frac{I_2^2I_3^2}{(2!)^2(3!)^2(1-I_1)^7}
+ 27\cdot \frac{1}{3!} \frac{I_3^3}{(3!)^3(1-I_1)^6} \\
& + & 240 \cdot \frac{1}{3!}\frac{I_2^3I_4}{(2!)^34!(1-I_1)^7}
+ 18 \cdot \frac{I_2I_3I_4}{2!3!4!(1-I_1)^6}\\
& + & \frac{36}{5} \cdot \frac{1}{2!} \frac{I_4^2}{(4!)^2(1-I_1)^5}
+ 40 \cdot \frac{1}{2!} \frac{I_2^2I_5}{(2!)^25!(1-I_1)^6} \\
& + & 8 \cdot \frac{I_2I_6}{2!6!(1-I_1)^5}
+ \frac{1}{8}\cdot 14 \cdot \frac{I_7}{7!(1-I_1)^4}
\biggr] +\cdots.
\een

\ben
 F_1(t)
& = &  t^2 \log \frac{1}{1-I_1}
+ t\biggl( \frac{1}{6} \cdot \frac{I_2^2}{2!^2(1-I_1)^3}
+  \frac{1}{4} \cdot \frac{I_3}{3!(1-I_1)^2} \biggr) \\
& + & t^2\biggl(\frac{1}{4!}\cdot 56
\cdot \frac{I_2^4}{(2!)^4(1-I_1)^6}
+ \frac{1}{2!}\cdot 13  \cdot \frac{I_2^2I_3}{(2!)^23!(1-I_1)^5} \\
&& + \frac{1}{2!} \cdot \frac{1}{4} \cdot 15  \cdot \frac{I_3^2}{(3!)^2(1-I_1)^4}
 + 4 \cdot \frac{I_2I_4}{2!4!(1-I_1)^4}
+ \frac{5}{3} \cdot \frac{I_5}{5!(1-I_1)^3} \biggr)
\een
plus
\ben
& + & t^3 \biggl[ 7104 \cdot \frac{1}{6!}\frac{I_2^6}{(2!)^6(1-I_1)^9}
+ 3648 \cdot \frac{1}{4!}\frac{I_2^4I_3}{(2!)^43!(1-I_1)^8} \\
& + & 384 \cdot \frac{1}{2!2!}\frac{I_2^2I_3^2}{(2!)^2(3!)^2(1-I_1)^7}
+ 99 \cdot \frac{1}{3!} \frac{I_3^3}{(3!)^3(1-I_1)^6} \\
& + & 612 \cdot \frac{1}{3!}\frac{I_2^3I_4}{(2!)^34!(1-I_1)^7}
+54  \cdot \frac{I_2I_3I_4}{2!3!4!(1-I_1)^6}\\
& + & 24 \cdot \frac{1}{2!} \frac{I_4^2}{(4!)^2(1-I_1)^5}
+ \frac{370}{3} \cdot \frac{1}{2!} \frac{I_2^2I_5}{(2!)^25!(1-I_1)^6} \\
& + & 30  \cdot \frac{I_2I_6}{2!6!(1-I_1)^5}
+ \frac{1}{8}\cdot 70 \cdot \frac{I_7}{7!(1-I_1)^4}
\biggr] +\cdots,
\een

\ben
  F_2(t)
& = &   t \biggl[  840  \cdot \frac{1}{6!}\frac{I_2^6}{(2!)^6(1-I_1)^9}
+ 630 \cdot \frac{1}{4!}\frac{I_2^4I_3}{(2!)^43!(1-I_1)^8} \\
& + & 70 \cdot \frac{1}{2!2!}\frac{I_2^2I_3^2}{(2!)^2(3!)^2(1-I_1)^7}
+\frac{45}{2}\cdot \frac{1}{3!} \frac{I_3^3}{(3!)^3(1-I_1)^6} \\
& + & 114t \cdot \frac{1}{3!}\frac{I_2^3I_4}{(2!)^34!(1-I_1)^7}
+ 12t \cdot \frac{I_2I_3I_4}{2!3!4!(1-I_1)^6}\\
& + & \frac{33}{5} \cdot \frac{1}{2!} \frac{I_4^2}{(4!)^2(1-I_1)^5}
+ 25 \cdot \frac{1}{2!} \frac{I_2^2I_5}{(2!)^25!(1-I_1)^6} \\
& + & 7t  \cdot \frac{I_2I_6}{2!6!(1-I_1)^5}
+ \frac{1}{8}\cdot 21  \cdot \frac{I_7}{7!(1-I_1)^4}
\biggr] +\cdots.
\een
One can also use the coordinates $\{q_n\}$ defined in \eqref{eqn:qn}
to simplify the expressions and to see a connection to partitions of even integers.
Such expression should be useful to the study of singular behavior of the free energy
and the (multi)critical phenomenon in Hermitian matrix models.
We hope to address such applications in the future investigations.

\vspace{.2in}
{\bf Acknowledgements}.
The author is partly supported by NSFC grant 11661131005.

\end{document}